\documentclass[nofigure]{pasj00}

\SetRunningHead{K. Wada}{Multi-phase gas in a bar}

\title{Multi-Phase Gas Dynamics in a Weak Barred Potential}
\author{Keiichi Wada}%
\affil{%
   National Astronomical Observatory of Japan, 2--21--1 Osawa,\\
   Mitaka, Tokyo 181--8588}
\email{wada.keiichi@nao.ac.jp}
\author{Jin Koda}%
\affil{%
   Institute of Astronomy, The University of Tokyo,\\
   Mitaka, Tokyo 181--8588}

\Received{$\langle$reception date$\rangle$}
\Accepted{$\langle$acception date$\rangle$}
\Published{$\langle$publication date$\rangle$}

\def\mbf#1{\mbox{\boldmath ${#1}$}}

\begin{document}
\maketitle

\begin{abstract}
The structure of the interstellar medium in 
the central kpc region of a galaxy with a weak bar-like potential 
is investigated taking into account realistic cooling and 
heating processes and 
the self-gravity of the gas.
Using high resolution hydrodynamical simulations, it is revealed that 
the resonant structures (e.g. smooth spiral shocks and a nuclear ring) 
are very different from those seen
in past numerical models where simple models of the ISM, i.e. 
non-self-gravitating, isothermal gas were assumed.
We find that the pc-scale filaments and clumps form large scale spirals,
which resemble those seen in real galaxies.
The fine structures are different between the arms and in 
the nuclear region. The next generation millimeter interferometer (ALMA) 
may reveal the fine structures of the cold gas in
nearby galaxies.

We also find a large scale anisotropy in the gas temperature, which is caused
due to non-circular velocity field of the gas. 
The damped orbit model based on the
epicyclic approximation explains the distribution of the hot ($> 10^4$ K) and cold ($ < 100$ K) gases appearing alternately around the galactic center.
Because of the temperature anisotropy,
cold gases observed by molecular lines do not necessarily represent
the real gas distribution in galaxies.

Position-Velocity diagrams depend strongly on the viewing angles.
As a result, 
the rotational velocity inferred from the PV maps
could be two times larger or smaller than the true circular velocity.

\end{abstract}

\section{INTRODUCTION}
\label{intro}
Gaseous response to the bi-symmetric gravitational 
potential, such as a stellar bar in galaxies,
has been well studied numerically 
since early 70's 
\citep{mil70,SM76,mat77,hun78,mat80,alb81,sch84,mat87,AT92,WH92,
fri93,WH95,pin95,HS94,EG97,FW98,fux99,pat00,eng00,fuk00}.
A rotating bar-like potential causes resonances in gas orbits, 
and spiral-like density enhancement or shocks are formed in the gas disk.
However, almost all simulations and analyses in the past three decades assumed
a simple model of the interstellar medium (ISM) with 
the isothermal equation of state, where the 
gaseous temperature is assumed  a $\sim 10^4$ K.
In some models, the energy equation is solved, 
but $10^4$ K cutoff is introduced \citep{fri93}.
In another major numerical representation of the ISM, i.e. the 
``sticky particle'' method or cloud-fluid model \citep{com85,fuk91},
the ISM particles loose their energy via inelastic collisions,
and the velocity dispersion of the particles are kept roughly constant.
The isothermal equation of state or constant velocity dispersion would be 
a relevant approximation to investigate the global gas dynamics in galaxies.
It is obvious, however, that the interstellar matter is not an isothermal and
uniform media \citep{MO77},
the inhomogeneous nature of the ISM should be 
taken into account to investigate gas dynamics in a local scale or 
on a galactic central region.
In the case that we have no observational information about 
the ISM with fine enough spatial resolution (e.g. parsec scale) 
in external galaxies, 
the simple treatment of the ISM is enough to compare the models with
observations. Hopefully we will have extremely high quality observations
on the ISM in the next decade.
High resolution observations with
the next generation millimeter/submillimeter interferometer (e.g. ALMA)
will reveal the molecular gas structure with a hundred 
times finer spatial resolution.
Therefore more realistic treatment of the ISM in simulations on a galactic 
scale is necessary to understand the ISM through comparison
 with the observations.

Recently Wada \& Norman (1999, 2001) 
have investigated dynamics and structure of a
self-gravitating gas disk in the central kpc region of a galaxy, taking into
account realistic cooling (10 K $< T_g < 10^8 $K) and heating.
Using high accuracy Euler mesh code,
they showed that the globally stable, 
inhomogenoeus interstellar matter is formed as natural consequences of 
gravitational and thermal instabilities. The surface density ranges 
from $10^{-2}$ to $10^{6} M_\odot$ pc$^{-2}$, and the
dense gas formes clumps or filaments where the temperature is 
$T_g \sim 10-100$ K. In the diffuse gas, on the other hand, the 
gaseous temperature is $T_g \sim 10^4 - 10^5$ K as a result of 
shock heating or $T_g \sim 10^6-10^7$ K due to supernova explosions.

In this paper, we invesigate the gas dynamics in a weak barred potential
using the same numerical method presented by \citet{WN01} (hereafter WN01).
Energy feedback from stars and supernovae, which are implimented in WN01,
are not considered here for simplicity (see Wada, Spaans, \& Kim 2001
for effects of the energy feedback to the ISM).

This paper is organized as follows.  In \S 2 we give a brief
description of our numerical code and models.  In \S 3.1 
numerical results of the global structure of the
self-gravitating multi-phase ISM in a weak barred potential
are discussed. Effects of the advective heating and 
cooling are analytically studied in \S 3.2.
In \S 4, observational implications derived from
the position velocity diagrams and 
rotation curves are discussed. 
Finally we summarize the results in \S 5.

\section{NUMERICAL METHODS AND MODELS}

In order to model the global evolution of the multi-phase ISM, a
hydrodynamical code must deal with a wide dynamic range (e.g., scale:
1 pc $\sim $ a few kpc, density: $10^{-3} \sim 10^6$ cm$^{-3}$, and
temperature: $10 \sim 10^8$ K).  Moreover, shocks must be resolved
correctly, because typical Mach numbers range from 10 to 100 in the
ISM of galaxies, and also SNe and stellar winds cause strong shocks
with Mach numbers of several hundreds.
We also need high spatial resolution for
the hot, diffuse gas as well as for the cold, dense gas.
In a rotating non-axisymmetric potential, the dynamical resonances play
an important role for producing spiral shocks, density waves, and 
ring-like strucutres. We are interested in how 
the local inhomogeneous structure of the ISM affects
formation of these global resonant structures.

We use hydrodynamical simulations for a single fluid taking into
account the self-gravity of the gas, radiative cooling, and heating
processes. The multi-phase feature of the gas, that is, the
multi-temperature and multi-density structure, should be achieved as a
result of the non-linear evolution of a single fluid, if we take into
account realistic cooling and heating processes, and the self-gravity
of the gas with relevant spatial resolution.  We investigate
the effects of the global dynamics of the gas in a bar potential, therefore
it requires that the whole rotating disk is simulated.
The hydrodynamical part of the simulations
is solved by a high-accuracy explicit Eulerian code with non-adaptive
numerical grids. The effect of the radiative cooling is implemented using an
implicit method with a cooling function.  The numerical code is tested
for various standard problems (see details WN01).


We solve the following equations numerically 
on a rotating frame of a bar potential 
in two dimensions.
\begin{eqnarray}
\frac{\partial \rho}{\partial t} + \nabla \cdot (\rho \mbf{v}) &=& 0, \label{eqn: rho} \label{eq1} \\ 
\frac{\partial \mbf{v}}{\partial t} + (\mbf{v} \cdot \nabla)\mbf{v}
+\frac{\nabla p}{\rho} &=& -\nabla \Phi_{\rm ext} - \nabla \Phi_{\rm sg} \\ \nonumber
&-& \mbf{\Omega}_b \times (\mbf{\Omega}_b \times \mbf{r}) 
 -2\mbf{\Omega}_b\times\mbf{v}
 , \label{eqn: rhov}  \label{eq2} \\  
\frac{\partial E}{\partial t} + \frac{1}{\rho} \nabla \cdot [(\rho E+p)\mbf{v}] &=& \Gamma_{\rm UV}- \rho \Lambda(T_g), \label{eq3}\\
\nabla^2 \Phi_{\rm sg} &=& 4 \pi G \rho, \label{eqn: poi}
\end{eqnarray}
where, $\rho,p,\mbf{v}$ are density, pressure, and velocity of the gas, 
the specific total energy $E \equiv |\mbf{v}|^2/2+ p/(\gamma -1)\rho$
with $\gamma = 5/3$, and $\Omega_b\equiv |\mbf{\Omega}_b|$
 is the pattern speed of the bar potential. 

The potential of the analytical models is
\begin{equation}
\Phi_{\rm ext}(R,\phi) = \Phi_0(R) + \Phi_1(R,\phi), 
\end{equation}
where $\Phi_0$ and $\Phi_1$ are axisymmetric and
non-axisymmetric potentials, respectively.
We use the `Toomre disk' potential for the 
axisymmetric part. 
The potential has two parameters $a$ (scale parameter called `core radius') and $v_{\rm max}$ (the maximum rotation speed), and expressed as
\begin{equation}
\Phi_0 (R)=  \frac{c^2}{a {(R^2 + a^2)}^{1/2} } \  ,  \label{eq-1}
\end{equation}
 where $c $ is given as  $c \equiv v_{\rm max}
{(27/4) }^{1/4} a $. 
The non-axisymmetric part of the potential is assumed to be
in the form
\begin{equation}
 \Phi_1 (R, \phi, t) =   \varepsilon (R) \Phi_0 \cos 2(\phi -\Omega_b t)    \ ,
\end{equation}
where $\varepsilon (R) $ is given as
\begin{equation}
 \varepsilon(R)= \varepsilon _0  \frac{aR^2}{(R^2+a^2)^{3/2}}.
\end{equation}
The parameter $\varepsilon_0$ represents the strength of the bar. 
This formalism 
is taken from \citet{san77}.
The parameter  $\varepsilon_0$ is at full strength from the beginning of
simulations. Since our bar models are weak ($\varepsilon_0 =$ 0.05, i.e. 
$\Phi_1/\Phi_0 (a) = 0.02$
), initial strong shocks and 
subsequent accretion of gas to the galactic center do not occur.

The equations (1), (2), and (3) are solved by second-order AUSM (Advection
Upstream Splitting Method) based on \citet{LS}.
We achieve third-order spatial accuracy with the Monotone
Upstream-Centered Schemes for Conservation Laws (MUSCL) approach
(\cite{VL}).

Time integration is achieved by the
second-order leap-frog method.  Each time step is
determined by the Courant condition.
We adopt implicit time integration for the cooling term. 

Various two-dimensional test problems, such as the double Mach
reflection of a strong shock, a point explosion in an adiabatic uniform
medium, an isothermal gas flow in a weak barred potential, and
evolution of spiral gravitational waves in a differentially rotating disk
are described in \citet{WN01}.
We find that AUSM with MUSCL is as powerful a scheme for
astrophysical problems as PPM (\cite{WC}), but
the algorithms and coding are much simpler than PPM.

We use $2048^2$ Cartesian grid points for calculating gas dynamics in
a 2 kpc$\times$ 2 kpc region which includes the galactic center.
Therefore the spatial resolution is 0.98 pc.
The gravitational potential of the gas 
is calculated using 
the convolution method with the Fast Fourier Transform (\cite{HE}).
In this method, $4096^2$ grid cells and a periodic Green's function
are used to calculated the self-gravity of the isolated
matter in the $2048^2$ grid points.

We also assume a cooling function $\Lambda(T_g) $ $(10 < T_g < 10^8 \, {\rm K})$ and the photoelectric heating rate $\Gamma_{\rm UV}$.
We use the cooling curves of Spaans \& Norman (1997) (Fig.1 in WN01).  
The cooling processes taken into account are,
(1) recombination of H, He, C, O, N, Si and Fe, (2)
collisional excitation of HI, CI-IV and OI-IV, (3) hydrogen and helium
bremsstrahlung, (4) vibrational and rotational excitation of H$_2$ and
(5) atomic and molecular cooling due to fine-structure emission of C,
C$^+$ and O, and rotational line emission of CO and H$_2$.
We assume solar metalicity in this paper.

We assume a
uniform UV radiation field and photoelectric heating of grains and
PAHs.  The heating rate $\Gamma_{\rm UV}$ is the same as in
\citet{ger97}: $ \Gamma_{\rm UV} = 1.0 \times 10^{-24} \varepsilon
G_0 \, {\rm ergs \: s}^{-1}, $
where the heating efficiency $\varepsilon$ is assumed to be 0.05 and
$G_0$ is the incident FUV field normalized to the local interstellar
value.

The initial condition is an axisymmetric, uniform and rotationally supported
disk, whose radius is 1 kpc, with the total gas 
mass $5\times 10^7 M_\odot$ which is  about 
2 \% of the total mass determined by the background gravitational potential.
The initial rotational velocity is chosen in order to balance the
centrifugal force caused by the external potential $\Phi_{\rm ext}$
and $\Phi_{\rm sg}$.

Random density fluctuations are added to the
initial disk, with an amplitude less than 1 \% of the unperturbed
density and temperature. The initial temperature is set to $10^4$ K
over the whole region. In ghost zones at boundaries, the physical
values remain at their initial values during the calculations. From
test runs, we found that these boundary conditions are much better than
so called `outflow' boundaries, because the latter cause strong
unphysical reflection of waves at the boundaries.

%
\section{NUMERICAl RESULTS}
%
\subsection{Non-isothermal, self-gravitating gas in a bar potential}
For comparison with the realistic model, 
we first show the density and temperature distribution
 of a model without taking into account 
self-gravity of the gas in Fig. 1, which is a snapshot at $t = 20$ Myr 
(about 2.4 rotational period at $R = 0.2$ kpc). The surface 
density map (Fig. 1(a))
shows a typical gas response to a weak bar potential with two inner
Lindblad resonances (IILR and OILR). The two inner {\it leading} spirals 
form an oval structure. 
The outer two-arm {\it trailing}
spiral is a consequence of the OILR \citep{wad94}.
The temperature map (Fig. 1 (b)) shows a quadruple distribution. 
The gas is heated up to $10^4 \sim 10^5$ K 
at upstream sides of the leading and 
the trailing spirals. At downstream regimes, the gas temperature is
as low as the lower limit of the temperature (10 K) to the contrary.
This extreme anisotropy is because the advective cooling/heating that
dominates the heat balance in the supersonic non-axisymmetric flow.
See details in \S 3.2. 

If we take into account self-gravity of the gas, however, the density and
temperature distributions 
in a quasi-stable state have very different fine structures from
Fig. 1. Figure 2 shows
density and temperature maps of the self-gravitating model at $t=61$ Myr.
As seen in the density map (Fig. 2(a)), the clumpy and filamentary
dense gases form weak spiral density enhancements, whose location roughly
 coincides with those in the non-self-gravitating model.
The cold gases form complexes of clouds in the spirals
that look similar to the giant molecular clouds in spiral galaxies.

The complicated network of the filaments and clumps on a local scale 
is the same as that in axisymmetric models in WN01.
A number of processes are involved in the
formation of the filamentary structure: 1) tidal and collisional 
interactions between dense regions formed due to the 
thermal and gravitational instabilities, 2) differential rotation, and 
3) shear motion due to local turbulent motion, 

The closeup of the central 500 pc region reaveals the fine structure of the
gas (Fig. 3(a)).
The dense clumps in the nuclear region are typically 
20-50 pc across, and their masses $\sim 10^5 M_\odot$. The filamentary 
structures, which are nearly parallel to the bar major axis,
 are prominent.
They are formed due to non-axisymmetric motion and shear in this region.
The density range is greater than four orders of magnitude.
On other hand, the closeup of the arm region (Fig. 3(b)) shows that
its fine structure is very different from that in the nuclear region. The size
of the dense clumps in the arms are smaller than that in the central region,
because the velocity dispersion is smaller in the arm regions.
This result 
suggests that the star formation processes in the arm and in the central
regions would be different. 

Comparing the realistic ISM model (Fig. 2) with the non-selfgravitating
 model (Fig. 1),
the continuous galactic shocks seen
in the non-selfgravitating model,
is no longer seen in the realistic model.
Spiral structures are actually ensembles of dense clouds and filaments.
 The continuous central 
ring seen in Fig. 1 is not also 
reproduced in the self-gravitating model, where the central density
structure is very complicated as seen in Fig. 3 (a). 
These results suggest that
self-gravity of the gas should not be negligible 
even if the gas mass fraction to the dynamical 
mass is small (2 \% in the present case), in order to
explore the gas dynamics in the central kpc region of galaxies.
The isothermal approximation for the ISM, which was often used in 
past numerical simulations, would be relevant for a galactic scale 
gas dynamics, but more realistic treatment of the cooling 
processes should be necessary for the ISM on a smaller scale. 
We should take into account
 small scale structure of the ISM especially for the cases 
that star formation and its energy feedback to the ISM are important.
If one uses the isothermal gas model for the ISM with star formation, 
it is hard to avoid to introduce some phenomenological models for 
the star formation and energy feedback processes.  
Our numerical approach, which is based on
the basic equations that represent the phenomena,
is less ambiguous in terms of these points.

\subsection{Effect of the advective cooling in a non-uniform stream
motion in a bar potential}

We mentioned in the previous section that temperature of the gas
in a bar potential changes quadruply on a global scale,
 i.e. cold ($T_g \lesssim 100$ K) and hot ($T_g \gtrsim 10^4$ K)
regimes appear alternately around the galactic center (Fig. 1(b) and 2(b)).
This is because the advective cooling/heating dominates the heat balance 
in this system. This can be explained analytically 
using the ``damped gas orbit model'' under the epicycle approximation.

\citet{wad94} provided an analytical representation
 which represents the behaviour of 
a non-self-gravitating
gas in a rotating bar-like potential. 
Its outline is as follows.
The gas dynamics can be understood in terms of closed elliptical
orbits which are solutions to the equations of motion of a forced-damping
oscillator driven by a periodic external force.
Gaseous elliptical orbits always incline to the bar 
potential in a leading sense inside of the corotation, and in a following sense 
outside of the corotation.
If there are two ILRs, the gas 
orbits at the ILRs are oriented by 45$^\circ$ with respect to the bar potential.
The damped orbits gradually change their orientation with 
their radius.
Since the direction of the gradual rotation near to the first ILR and
that near to the second ILR are in an opposite sense, leading or trailing
spiral-like enhancements appear around the first and the second ILR,
respectively. 
This gradual rotation is explained in terms of the
phase-shift of a damping oscillator.
The analytical model well describes the gas behaviour seen 
in hydrodynamical simulations.

Under the epicycle approximation,
the simplest way to represent the effect of the gaseous nature, such as 
viscosity, on 
closed  
orbits is to introduce a simple damping term,
which is proportional to the velocity of the radial 
oscillation $2 \lambda \, \dot R_1$ into the equation of motion.
\begin{eqnarray}
\ddot R_1+2\lambda \,\dot R_1+\kappa _0^2R_1=f_0
\cos 2(\Omega_0 - \Omega_{\rm b}) \,t, \label{e10}
\end{eqnarray}
where
\begin{eqnarray}
f_0 \equiv -\left[ {{{d\Phi _{\rm b}} \over {dR}}+{{2\Omega \Phi _{\rm b}} 
\over {R(\Omega -\Omega _{\rm b})}}} \right]_{R_0} \, , \label{e11}
\end{eqnarray}
and $\lambda$ is the damping rate. 
This equation describes a radial damped oscillation at around $R_0$, 
which is driven by an external periodic force
at a frequency of $2(\Omega_0 - \Omega_{\rm b})$. The general 
solution to this equation is
\begin{eqnarray}
R_1(t)=A e^{-\lambda \kern 1ptt}\cos (\omega \kern 1ptt+\alpha )
+ B  \cos[ 2\left( {\Omega _0-\Omega_{\rm b}} \right)\,t+\delta_0 ] 
\, , \label{e12} 
\end{eqnarray}
where $A$ and $\alpha$ are arbitrary constants, and 
$ \omega \equiv \sqrt {\kappa _0^2-\lambda ^2 } $.
The first term of the right-hand side of (\ref{e12}) is a general 
solution of a
harmonic oscillator having a natural frequency of $\omega$ with weak 
friction.
Amplitude $B$ and the phase-shift, $\delta_0 \equiv \delta(R_0)$, are
\begin{eqnarray}
B \equiv {f_0 \over {\sqrt {\left\{ {\kappa _0^2-4(\Omega _0-\Omega 
_{\rm b})^2}
\right\}^2+16\lambda ^{\kern 1pt2}(\Omega _0-\Omega_{\rm b})^2}}}
\label{e13}
\end{eqnarray}
and
\begin{eqnarray}
\delta_0 =\arctan \left[ {{{2F\Theta} \over {F^2-1}}} \right] \, ,
\label{e14}
\end{eqnarray}
where
\begin{eqnarray}
F\equiv 2(\Omega _0-\Omega_{\rm b})/\kappa _0\;,\kern 1pt\;\Theta
\equiv \lambda /\kappa _0 \, .
\label{e15}
\end{eqnarray}
The phase-shift $\delta_0$ is 
always 
negative if $F > 0$, and $\delta_0 = -\pi/2$ when $F^2 = 1$, 
that is, at the Lindblad resonances ($\Omega_0 =
\Omega_{\rm b} \pm \kappa_0/2$). The negative $\delta_0$ means that the 
damping oscillation is delayed
from the periodic driving force. 
Since the first term on the right-hand side of equation (\ref{e12})
is a damping term which does not represent a closed orbit, we ignore it.
Using $ \phi _0 (t) =\left( {\Omega _0-\Omega_{\rm b}} \right)\,t $,
we obtain a closed orbit $(R_1,\phi_1)$, 
\begin{eqnarray}
R_1 (\phi_0) = B \cos \left( {2\phi_0 + \delta_0} \right) \, ,
\label{e16}
\end{eqnarray}
and
\begin{eqnarray}
\phi_1 (\phi_0) =  -\frac{\Omega_0 B}{R_0 (\Omega_0-\Omega_b)}
\left[ \sin(\delta_0) - \sin(2\phi_0 + \delta_0) \right]  \nonumber \\
- \frac{\Phi_b(R)}{2R_0^2 (\Omega_0 -\Omega_b)^2} \sin(2\phi_0). \label{e16b}
\end{eqnarray}
We plot the orbits for the bar potential,
\begin{eqnarray}
\Phi_{\rm bar} = (R^2 + b^2)^{-1/2} \left[1 + \varepsilon_0 b R^2 (R^2 +b^2)^{-3/2} \right]
\end{eqnarray}
 with
$\varepsilon_0 = 0.05, b=0.5, \Omega_b=0.25, \lambda=0.1$ in Fig. 4.

The gas velocity field ($v_R, v_\phi$) is obtained from (\ref{e16}) and (\ref{e16b}):

\begin{eqnarray}
v_R (R_0) &=& -B (\Omega_0 - \Omega_b) \sin(\phi_0 + \delta_0), \label{e18} \\
v_\phi (R_0,\phi_0) &=& [R_0 + B\cos(\phi_0+\delta_0)]  \nonumber \\
&\times& \left[
-\frac{2\Omega_0 B}{R_0} \cos(2\phi_0 +\delta_0) 
-  \frac{\Omega_b(R_0)}{R_0^2(\Omega_0-\Omega_b)} \cos(2\phi_0) 
\right]  \label{e19}
\end{eqnarray}


Now, the energy equation in the hydrodynamical simulations can be written as
\begin{eqnarray}
\frac{\partial \rho E}{\partial t} + \frac{\partial \rho H}{\partial x}
+ \frac{\partial \rho H}{\partial y} = \rho \Gamma_{\rm UV} - \rho^2 \Lambda(T_g)  \label{eq20}
\end{eqnarray}
where specific enthalpy $H \equiv E + p/\rho$ with $E = |\mbf{v}|^2+p/(\gamma -1)\rho$.

The advective cooling/heating rate is about one order of magnitude
larger than the radiative cooling/heating rate in the supersonic 
gas disk that we are exploring.
If the radiative cooling and heating is in an equilibrium (i.e.,
the right-hand side of eq. (\ref{eq20})), temperature change in a unit time 
due to the advective terms can be written as
\begin{eqnarray}
 \Delta T_{\rm ad}(R,\phi) &\sim& -\frac{\mu m_H (\gamma -1)}{k_B}
 \left[\frac{\partial |\mbf{v}|^2}{\partial R}
+ \frac{\partial |\mbf{v}|^2}{R\partial \phi} \right], \label{eq100} \\
&\sim & 1500 \left| \frac{\partial |\mbf{v}|^2}{\partial R}
+ \frac{\partial |\mbf{v}|^2}{R\partial \phi} \right| {\rm K}\;
{\rm Myr}^{-1}, \label{eq101} 
\end{eqnarray}
where $v^2 \gg c_s^2$ and $\rho \sim const$ is assumed for simplicity.
Using (18) and (19), we plot $-\left[{\partial |\mbf{v}|^2}/{\partial R}
+ {\partial |\mbf{v}|^2}/{R\partial \phi} \right] $ in Fig. 5.
$\Delta T_{\rm ad}$ is about $\pm 10^3 -10^4$ K Myr$^{-1}$.
The quadruple anisotropy of the advective cooling and heating 
in the hydrodynamical simulations (Fig. 1 and 2) is well
reproduced, and it suggests that the non-axisymmetric velocity field of
the gas dominates the global temperature distribution in a weak barred
potential.

\section{OBSERVATIONAL IMPLICATIONS}
From the numerical results shown in \S 3.1, we can infer how the 
molecular gas in nearby galaxies is observed with
the submillimeter interferometers. 
It is crucial to treat realistically the cooling and heating processes
of the ISM, especially when one tries to compare the numerical result
with observations, such as molecular line observations.
As shown in \S 3.1, the temperatures of the ISM in 
a bar potential globally anisotropic. 
This is essential to interpret the observed molecular line intensity
distribution. To make detail comparison between the models and observations,
one needs radiative transfer calculations for molecular lines (see 
Wada, Spaans, \& Kim 2000 for the Large Magellanic Clouds).
Alternatively, here we can use the density of 
the cold gas ($T_g < 100$ K) component instead of the line intensity
derived from detail line transfer calculations in order to infer
the observed molecular gas distribution. 

Figure 6 (a) is the 
surface density distribution of the cold gas ($<100 $ K) 
with a 100 pc resolution.
One may regard this map as a ``CO integrated intensity map'' of a galaxy 
in the Virgo cluster observed with the 
Nobeyama Millimeter Interferometer, for example.
Two clumpy trailing spirals are clearly seen,
 and one might see a nuclear ``pseudo ring''.
There are dense regions at the base of the spirals and also 
at the nuclear region.
Figure 6 (b) is the same as Fig. 6 (a), but with 10 pc resolution.
It is clear that spirals are actually 
an ensemble of small clumps and filaments (see also Fig. 2(b)).
Comparison of the two `observations' suggests
 that spatial resolution of the 
present-day interferometers are still too coarse to discuss 
small scale (e.g. 100 pc) structure of the cold gas in nearby galaxies.
In other words, 
one should not be confused by the small scale inhomogeneity, which are
comparable to the beam size, seen in observed intensity maps.

The other important implication from our numerical results
is that distribution of the cold gas derived 
from molecular line intensity (e.g. CO (1-0)) 
does not necessarily represent
true mass distribution of the gas even on a galactic scale.
The advective cooling and
heating of the gas due to the non-circular 
 velocity field in a bar potential
dominate global distribution of the gaseous temperature. 
Suppose there is a uniform gas ring
between the IILR and OILR. The temperature of the gas ring is roughly divided 
into two hot ($T_g > 10^4$ K) and two cold ($T_g < 100$ K) regions.
Therefore we cannot observe gases in 
the hot regions with the CO (1-0) lines, for example,
and non-axisymmetric distribution of the gas is apparently observed.
A ``hidden'' hot gas component coexists with 
the nuclear ``pseudo-ring'' seen in Fig. 6 (a).


In order to resemble the observed position velocity
(PV) maps for the molecular gas, we have convolved the numerical
result (Fig. 2) with different `beam sizes' and velocity 
resolutions (Fig. 7 and 8).
The PV maps obtained for three different viewing angles are shown.
From comparison between the three figures, one can find that
the PV map strongly depends on the viewing angles ($\theta_v$). 
Along the bar major axis (i.e. $\theta_v = 0^\circ$), about a two times larger 
line-of-sight velocity than the {\it true} circular rotational velocity (the dashed
curve) is apparently observed. If we observe the same model
along the minor axis of the bar (i.e. $\theta_v =  90^\circ$), 
the PV diagram represents nearly rigid rotation, and 
the apparent circular velocity is about a half of the real circular velocity
at the core radius ($R \sim 200$ pc). 
Therefore if one observes the ISM in a edge-on galaxy that has a weak bar,
and evaluate the dynamical mass from the rotation curve,
then the mass could be four times larger or smaller than the real mass,
depending on the viewing angle.
This viewing angle effect originating in the non-circular gas motion
has been also known in previous isothermal simulations \citep{WT94,ath99}.
More quantitative analysis of the effect of non-circular motion on
the rotation curves is discussed in \citet{kod01}.

Figure 8 is the same as Fig. 7, but with low spatial and 
velocity resolution (100 pc and 5 km s$^{-1}$) which 
resembles the observations with millimeter interferometers such as Nobeyama
Milliliter Array. Although the surface density map does not 
reveal the fine structure of the cold gas, the low resolution 
PV map represents basically the same kinematical feature of the system.
The most notable difference between the high and low resolution PV maps is 
that there are high `intensity', 
steep velocity components ($\Delta v \sim 50-100$ km s$^{-1}$) in
the high resolution map. These features are caused by the {\it local} 
non-circular motion. Most prominent examples are seen in
the central region ($R< 100$ pc) in 
the PV map ($\theta_v = 0^\circ$ and $45^\circ$) of Fig. 7.
These features are originated from 
filamentary structures along the bar major axis seen in 
the density map.
Such features cannot be resolved as discrete components in the
low resolution PV map (Fig. 8).
Therefore, in the low resolution PV maps, 
one might interpret the rotation curve steeply rises near the
galactic center. 
One should note that even the bar potential assumed here is very weak
($\Phi_1/\Phi_0 < 0.02$), the non-circular motion of the gas is 
large enough to cause the viewing angle effect on 
the PV diagrams as seen in Fig. 7 and 8.

%
\section{CONCLUSIONS}
%

The parsec-scale structure of the interstellar medium (ISM) in the central kpc 
region of 
a galaxy with a weak bar-like potential 
is discussed based on our two dimensional, 
high resolution hydrodynamical simulations, taking into account realistic
cooling and self-gravity of the gas.
The results and conclusions are summarized as follows.
\begin{itemize}
\item Taking into account realistic cooling/heating processes and 
self-gravity of the gas, the smooth spiral shocks which appear in the non-self-gravitating, isothermal gas model are not formed in the ISM.
\item The resonant structures due to the inner Lindblad resonances,
 such as spirals, have actually parsec-scale 
fine structures, which are filamentary and clumpy dense gases (\S 3.1; Fig. 2).
The fine structures are different between the arms and 
the nuclear region (\S 3.1; Fig. 3)
\item There is large scale anisotropy of the 
gaseous temperature, which is 
due to the non-circular velocity field of the gas (\S 3.1; Fig. 1 and 2). 
The damped orbit model based on the
epicycle approximation explains the quadratic distribution of the hot ($> 10^4$ K) and cold ($ < 100$ K) gases around the galactic center (Fig. 4 and 5).
\item Because of the temperature anisotropy, cold gases observed by molecular lines do not necessarily represent all gas components even on a 
global scale (\S 3.2; Fig. 6).
\item Position-Velocity (PV) diagrams depend strongly on the viewing angles 
even if the bar component is very weak.
The line-of sight velocity inferred from the PV maps 
along the major axis of the bar potential 
is about four times larger than that along the minor axis (\S 4; Fig. 7 and 8).
\item PV diagrams obtained using the same 
spatial and velocity resolutions in the present millimeter interferometers
globally show the same kinematics of the gas with
ten times finer resolutions. However, the low resolution PV maps do not
reveal the fine velocity structure, and as a result, steeply rising 
rotation curves near the galactic center ($R \lesssim 100$ pc) can be
apparently observed (\S 4; Fig. 7 and 8).
\item The spatial resolution of the present millimeter interferometers for external galaxies are not fine enough to discuss the gas dynamics on
a hundred parsec scale, and we should wait for
the next generation radio interferometers, e.g. ALMA (Fig. 6).
\end{itemize}

Most numerical simulations
on the gas dynamics in a non-axisymmetric
gravitational potential  in the last three decades 
have assumed the simple model of the ISM. Under such approximation,
the gas dynamics is well understood theoretically. 
However, we should investigate again 
the dynamics and structure of the ISM in galaxies 
using more realistic model based on the relevant basic equations.

\vspace*{0.5cm}

We are grateful to M. Spaans for providing us the cooling functions.
Numerical computations were carried out on Fujitsu VPP5000 at
the Astronomical Data Analysis Center of the National Astronomical
Observatory, Japan.

\newpage
\begin{figure}[ht]
 \caption{(a) Density distribution of a model without self-gravity of the
gas. The major-axis of the weak bar potential of a pattern speed,
$\Omega_b = 140$ km s$^{-1}$ kpc$^{-2}$
is horizontally fixed. 512$\times$ 512 grid points are used. Color represents log-scaled density between 1 and $10^3 M_\odot$ pc$^{-2}$. Gas rotates counter clockwise. Inner Lindblad Resonances are rocated at $R = $ 0.2 and 0.4 kpc. (b) Same as (a), but for temperature between 10 and $10^5$ K. }
\end{figure}

\begin{figure}[ht]
\caption{Same as Fig. 1, but self-gravity of the gas is taking into account.
2048$\times$2048 grid points are used.}
\end{figure}

\begin{figure}[ht]
\caption{ (a) Close-up of the central 500$\times$ 500 pc region in the 
same model of Fig. 2. The color scale represents log-scaled density 
between 0.1 and $10^3 M_\odot$ pc$^{-2}$.
(b) Arm-region}
\end{figure}

\begin{figure}[ht]
\caption{Gas orbits under the influence of the two ILRs, 
derived from the damped orbit model (eqn. (15) and (16)).} 
\end{figure}

\begin{figure}[ht]
\caption{Distribution of adiabatic cooling/heating rate, 
$-\left[{\partial |\mbf{v}|^2}/{\partial R}
+ {\partial |\mbf{v}|^2}/{R\partial \phi} \right] $. See eqn. (22).
Contours are drawn every 1 from -10 to 10. Solid/dotted lines represent
heating/cooling regions.}
\end{figure}

\begin{figure}[ht]
\caption{(a) Density distribution of the cold gas ($T_g < 100$ K) 
in a weak bar potential convolved with
a 100 pc ``beam'' size. (b) Same as left, but with a 10 pc beam size. 
Unit of scale is kpc.}
\end{figure}

\begin{figure}[ht]
\end{figure}

\begin{figure}[ht]
\caption{Surface density distribution of the cold gas 
with 10 pc grid size and 1 km s$^{-1}$ for velocity resolution.
Position-velcity diagrams from three viewing angles (0, 45, and
90 degree from the bar major axis) are plotted. The dashed curve
is a rotation curve derived from the axisymmetric gravitational potential 
eq. (\ref{eq-1}).}
\end{figure}

\begin{figure}[ht]
\caption{Same as Fig. 7, but with 100 pc and 5 km s$^{-1}$ 
spatial and velocity resolution. }
\end{figure}

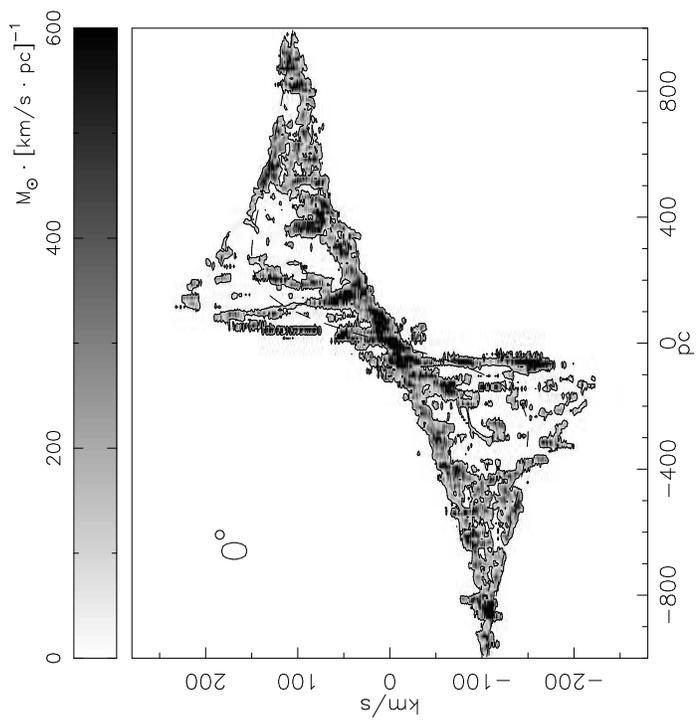
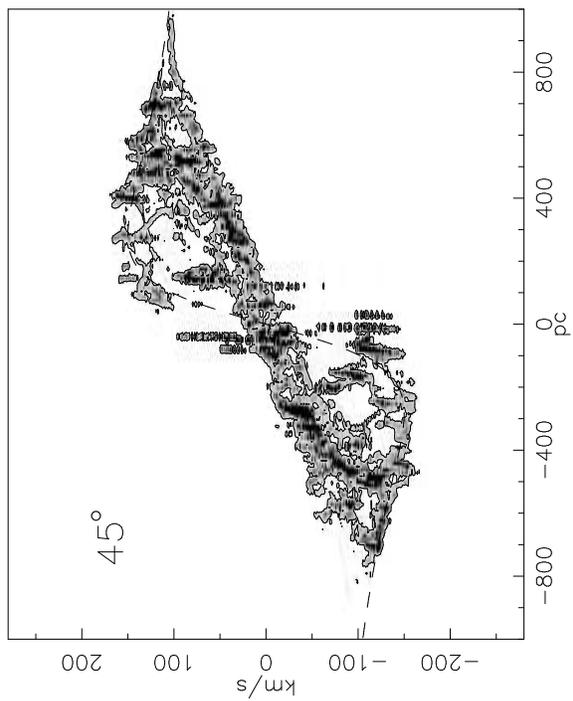
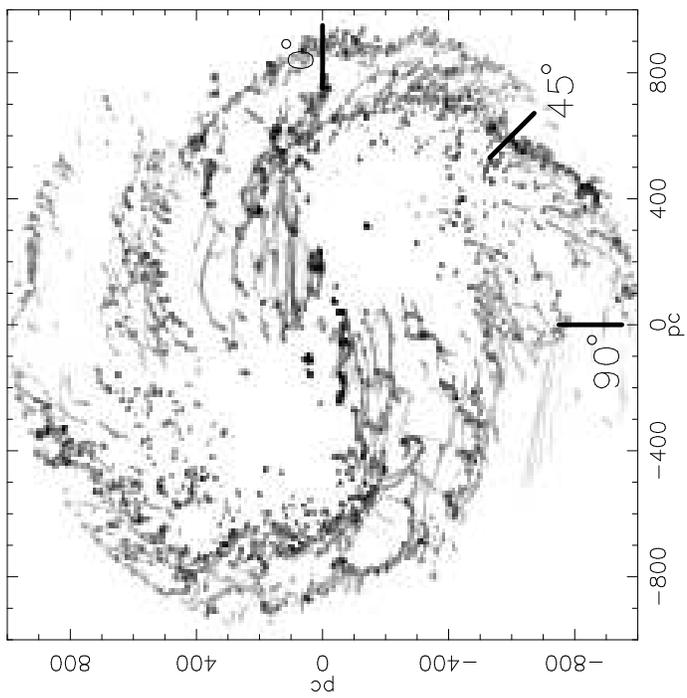
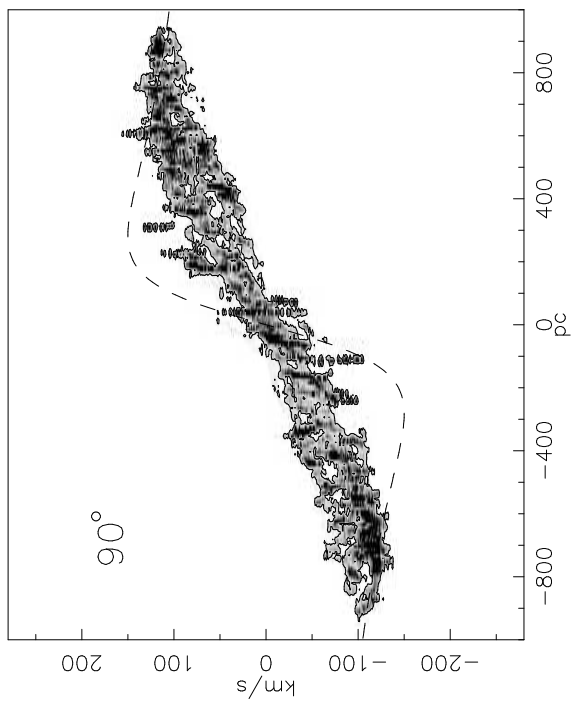
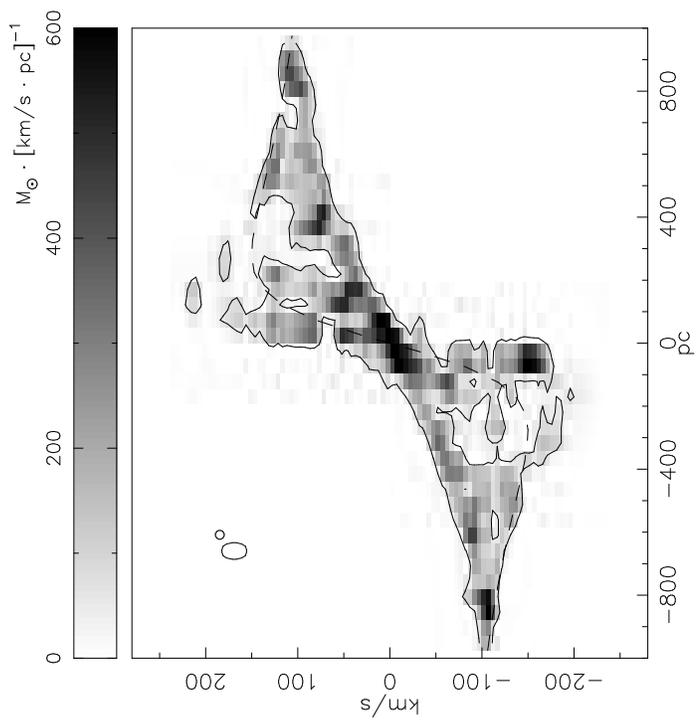
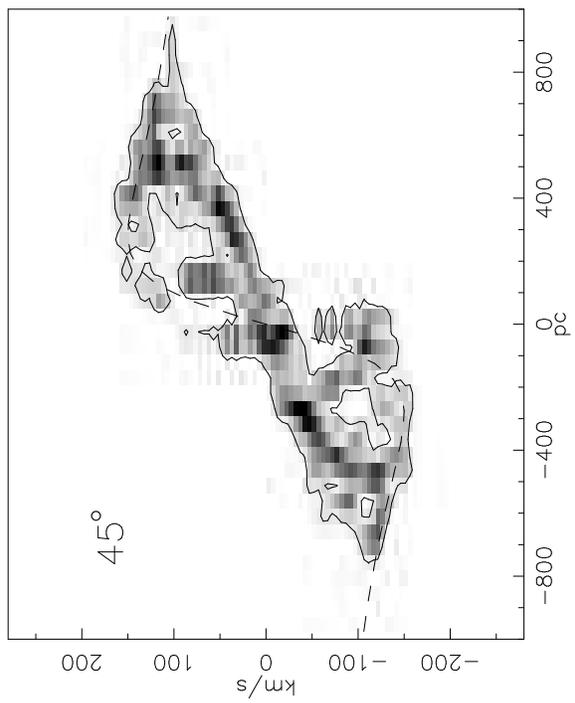
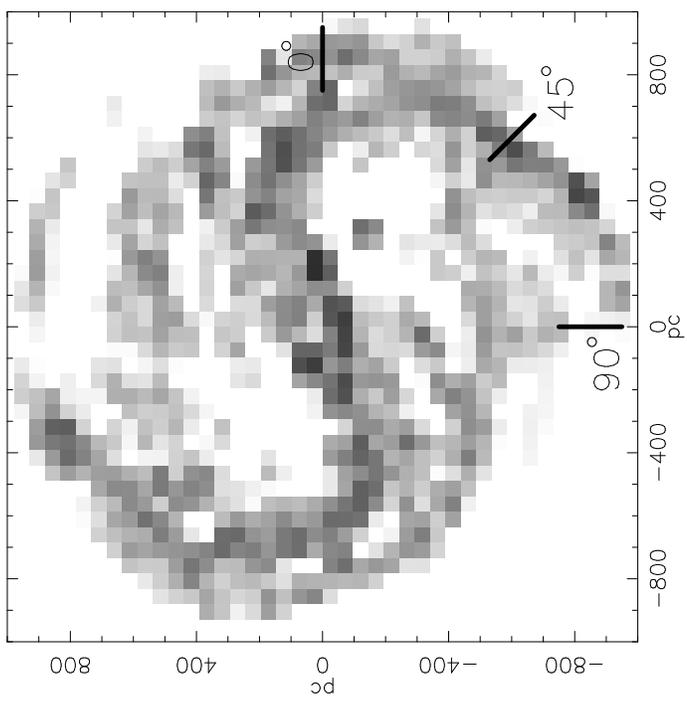
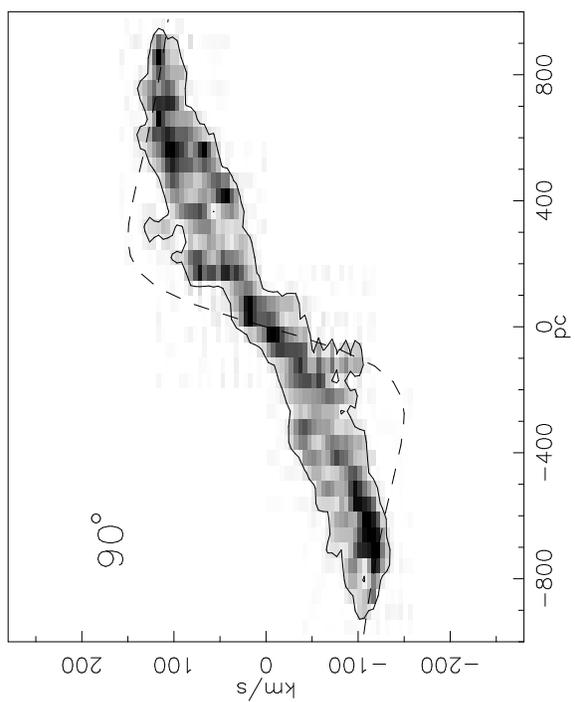
\end{document}